\documentclass[prb,preprint]{revtex4}

\usepackage{graphicx}
\begin{document}

\title{Observation of magnetic excitations of skyrmion crystal in a helimagnetic insulator}

\author{Y. Onose$^{1,2}$, Y. Okamura$^1$, S. Seki$^1$, S. Ishiwata$^1$, and Y. Tokura$^{1,3}$} 
\affiliation{$^1$Department of Applied Physics and Quantum Phase Electronics Center (QPEC),
University of Tokyo, Tokyo 113-8656, Japan
\\$^2$Department of Basic Science, University of Tokyo, Tokyo, 153-8902, Japan. 
\\$^3$ Cross-Correlated Materials Research Group (CMRG)
and Correlated Electron Research Group (CERG),
RIKEN Advanced Science Institute, Wako 351-0198, Japan}


\begin{abstract}
\bf{The skyrmion is a topologically stable spin texture, in which the spin direction wraps a sphere\cite{skyrme}. The topological nature gives rise to emergent electromagnetic phenomena such as topological Hall effect\cite{mlee,neubauer,kanazawa}. Recently, the crystallization of nanoscale skyrmions was observed in chiral helimagnets with use of the neutron diffraction and Lorentz transmission electron microscopy\cite{yu,muhlbauer}. 
The skyrmions are quite mobile under electric current density as low as $\sim$ 10$^6$ A/m$^2$, and, in some cases, stable up to near room temperature\cite{yu2,jonietz}. 
These features suggest that the skyrmions may work as a magnetic stable variable similar to the bubble memory\cite{bubble} but equipped with topological functionalities. 
Here, we investigate the low-energy excitations of the skyrmion crystal in a helimagnetic insulator Cu$_2$OSeO$_3$\cite{seki} in terms of microwave response. We have observed two distinct excitations of the skyrmion with different polarization characteristics; the counter-clockwise circulating mode at 1 GHz and the breathing mode at 1.5 GHz. These modes may play a crucial role in the low energy dynamics of skyrmions and hence in their manipulation via external stimuli.    
}
\end{abstract}
\maketitle
The manipulation of nano- or micro-scale magnetic objects has been the central issue of spin-related electronic science and technology. 
The bubble domains in the ferromagnets with uniaxial magnetic anisotropy can easily be manipulated by magnetic field gradient, and therefore were applied to the memory storage devise\cite{bubble}. The electrical control of ferromagnetic domain wall is thought to be a key technology for the magnetic random access memory\cite{maekawa}. 
Recently, a topological magnetic object termed skyrmion has been attracting much attention since the triangular lattice of skyrmions was observed in chiral helimagnets such as MnSi and (Fe,Co)Si\cite{muhlbauer,yu}.  
As shown in Fig. 1c, the magnetic moments composing the skyrmion point downward (opposite to the magnetic field) at the core and upward on the edge, and are aligned circularly in between.
It may be viewed as a nanoscale version of bubble memory but has a unique topological nature. 
Because the direction of magnetic moment varies continuously and wraps all the solid angle in the skyrmion, the skyrmion carries a topological quantum number termed skyrmion number
\begin{eqnarray}
N_s=\frac{1}{4\pi} \int d^2 {\bf r} \ {\bf n} \cdot (\nabla_x {\bf n} \times  \nabla_y  {\bf n})=-1,  
\end{eqnarray}     
where ${\bf n}$ is the unit vector parallel to the magnetic moment at the position ${\bf r}=(x,y,z)$. Reflecting the topological nature, the skyrmion works as a fictitious magnetic flux quantum $\phi_0=h/e$ and gives rise to emergent electromagnetic phenomena such as topological Hall effect, spin-motive force, and skyrmion Hall effect\cite{mlee,neubauer,kanazawa,schultz,zang}. 

The important issue to be clarified is the dynamics of skyrmions. It has recently been demonstrated that the skyrmion can be easily driven by electric current\cite{jonietz,schultz}. Jonietz {\it et al.} reported that the rotation of diffraction pattern is caused by the electrical current in the presence of the thermal gradient and ascribed it to the result of the current drive of skyrmions\cite{jonietz}. On the other hand, Schultz {\it et al.} have observed the emergent electric field or spin motive force induced by the skyrmion current\cite{schultz}. The threshold electric current ($\sim 10^6$ A/m$^2$) is five orders of magnitude lower than that needed to drive the ferromagnetic domain wall, which implies the potential of practical application. 

The resonant excitation by microwave irradiation provides a useful probe for dynamics of magnetic objects. In fact, the magnetic bubbles can be created and detected with use of the microwave\cite{bubble}. 
The oscillation of magnetic domain walls can also be excited by the AC magnetic field with the frequency of sub-GHz range\cite{osci}.   
Here, we have investigated the magnetic excitations of skyrmion crystal with use of microwave in a helimagnetic insulator Cu$_2$OSeO$_3$ with small Gilbert damping. 

\begin{figure}[htbp]
\begin{center}
\includegraphics*[width=15cm]{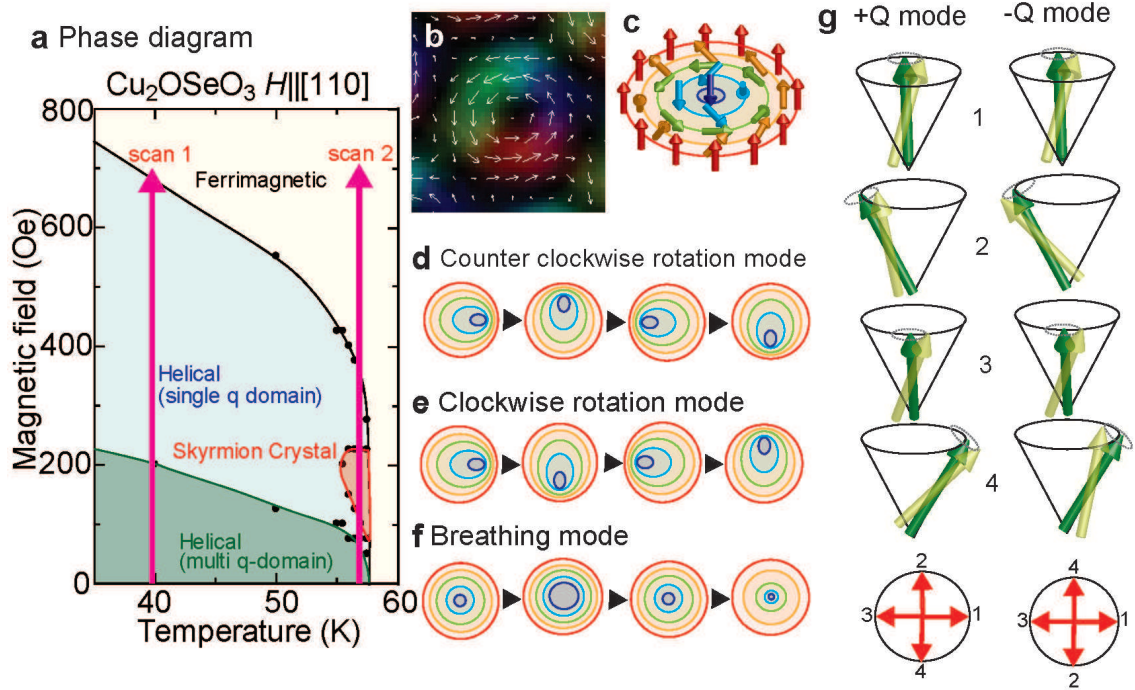}
\caption{{\bf Magnetic excitations in skyrmion crystal and helimagnetic state.} {\bf a,} Magnetic phase diagram for Cu$_2$OSeO$_3$ as determined by the magnetic susceptibility, which is measured with minimal demagnetization field (corresponding to the $H_{\rm DC} || H_{\rm AC}$ configuration in the following microwave experiments). {\bf b,} Real-space image of a skyrmion in Cu$_2$OSeO$_3$ obtained by Lorentz transmission electron microscopy, reproduced from ref. 10. {\bf c-f,} Illustrations of skyrmion ({\bf c}) and magnetic excitations in skyrmion crystal ({\bf d-f}). {\bf g.,} Illustrations of magnetic excitations in helimagentic state. The untransparent and half transparent magnetic moments indicate the magnetic moments in static and excited states, respectively.  
Bottom panel illustrates the oscillating component (the difference between the excited and static states) of magnetic moments in the coordinate system fixed to the static magnetic moment. The numbers stand for the positions of magnetic moments. }
\end{center}
\end{figure}

While the skyrmion crystal has been observed so far mostly in the B20 transition metal compounds such as MnSi and (Fe,Co)Si, it has recently been found that it can be realized also in an insulating oxide Cu$_2$OSeO$_3$\cite{seki}. In common with the B20 transition metal compounds, the space group of the crystal structure is $P2_13$, which is  noncentrosymmetric but nonpolar and cubic. Figure 1a shows the magnetic phase diagram for Cu$_2$OSeO$_3$\cite{seki}. 
The $S=1/2$ moments at the Cu$^{2+}$ sites show the local ferrimagentic arrangement of three-up and one-down type below $T_{\rm c} \sim 60$ K as observed by power neutron diffraction\cite{bos} and NMR\cite{belesi} measurements. However, the Dzyaloshinskii-Moriya interaction arising from the noncentrosymmetric lattice structure modulates the ferrimagnetic correlation and induces the helical spin structure with the period of about 50 nm at zero magnetic field. While the magnitude of the helical wave vector is determined by the ratio between the asymmetric Dzyaloshinskii-Moriya and symmetric ferrimagentic interactions, the wave vector direction is nearly degenerate, and the multi-domain structure is formed at zero magnetic field\cite{seki}. When the magnetic field is applied, the wave vector tends to be aligned along the field direction, forming the single-domain conical spin structure. In a high enough magnetic field, the induced ferrimagnetic state is realized. Just below the critical temperature, on the other hand, there appears the nontrivial magnetic state, {\it i.e.} skyrmion crystal\cite{seki}.  
The real space image of skyrmion as obtained by Lorentz transmission electron microscopy is reproduced in Fig. 1b, where the in-plane components of magnetic moments in the skyrmion are well resolved.

\begin{figure}[htbp]
\begin{center}
\includegraphics*[width=15cm]{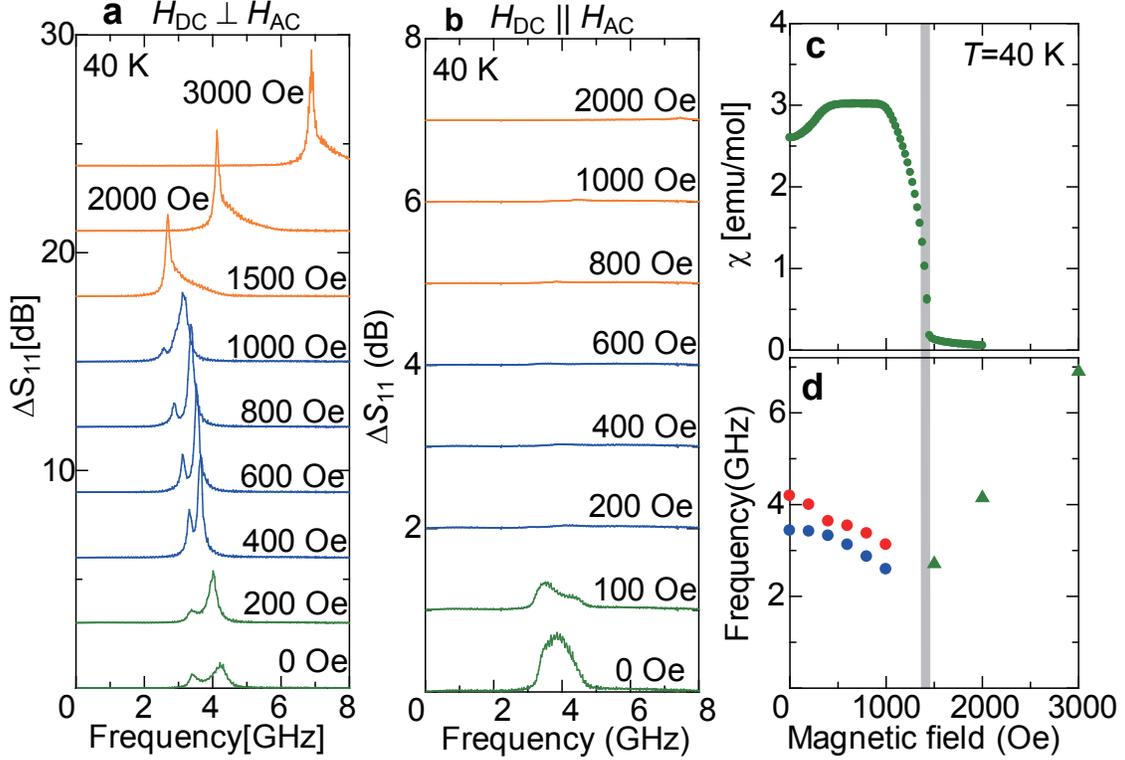}
\caption{{\bf Magnetic field dependence of microwave absorption spectra at 40 K. a,b,} Microwave absorption spectra $\Delta S_{11}$ at various magnetic fields for $H_{\rm DC} \perp H_{\rm AC}$ ({\bf a}) and $H_{\rm DC} || H_{\rm AC}$ ({\bf b}). {\bf c.,} The magnetic field variation of magnetic susceptibility at 40 K measured in the configuration corresponding to the $H_{\rm DC} \perp H_{\rm AC}$ microwave measurement. {\bf d} The magnetic field dependence of the resonance peak frequency in the microwave wave absorption spectra for $H_{\rm DC} \perp H_{\rm AC}$. } 
\end{center}
\end{figure}

The skyrmion crystal in a bulk crystal is restricted to a narrow range of magnetic field, therefore the microwave response can hardly be observed in a conventional ESR spectrometer, in which the probe frequency is fixed and the magnetic resonance is probed by scanning magnetic field. Thus, we have constructed a broad-band microwave system with a network analyzer as detailed in the Supplementary Information. While the pioneering work of electron spin resonance for MnSi was previously done with use of several fixed frequencies\cite{date}, our broad band measurement could provide much more information as shown below.  
At first, we show the microwave response at 40 K to discuss the magnetic excitations in the helimagnetic state (Scan 1 shown in Fig. 1a). 
In Figs. 2a and b, we show the magnetic field dependence of microwave absorption spectra $\Delta S_{11}$ at 40 K. 
(Here, $\Delta S_{11}$ stands for the difference of the reflection coefficient $S_{11}$ of the sample-inserted transmission line from the background value, which is proportional to the microwave absorption of the sample; for details, see Supplementary Information). The AC magnetic field of microwave $H_{\rm AC}$ is perpendicular to the DC magnetic field $H_{\rm DC}$ in a, and parallel in b. 
For the both cases, $H_{\rm DC}$ is applied along $<110>$ direction. The microwave spectra are qualitatively unchanged by the variation of $H_{\rm DC}$ direction with respect to the crystal axis although they are affected more or less by the static demagnetization field since we used a plane-like sample in order to fit it to the microwave probe. In the configuration of $H_{\rm DC} || H_{\rm AC}$, $H_{\rm DC}$ is parallel to the sample plane and the demagnetization field is small. On the other hand, $H_{\rm DC}$ is perpendicular to the sample plane and the demagnetization field is large in the $H_{\rm DC} \perp H_{\rm AC}$ configuration. 
$\Delta S_{11}$ shows broad two peaks around 4 GHz for both $H_{\rm DC} || H_{\rm AC}$ and $H_{\rm DC} \perp H_{\rm AC}$ in the low $H_{\rm DC}$ region. The peaks increase in intensity and     
become sharper when $H_{\rm DC}=400$ Oe is applied perpendicular to $H_{\rm AC}$, whereas they are completely suppressed above 200 Oe in the parallel configuration. 
As shown in Fig. 1a, the magnetic wave vector $q$ is aligned along $H_{\rm DC}$ forming a single $q$-domain above 200 Oe. (For the $H_{\rm DC} \perp H_{\rm AC}$ configuration, the critical magnetic field is $\sim $ 400 Oe becasue of the large demagnetization field.) 
Therefore, the contrastive polarization dependence in the high magnetic field indicates that the two magnetic modes can be excited only by $H_{\rm AC}$ perpendicular to the helimagnetic wave vector. Kataoka theoretically identified such two modes in the helical magnetic state induced by Dzyaloshinskii-Moriya interaction\cite{kataoka}. The two modes, +Q mode and -Q mode, are shown in Fig.  1g. In the coordinate fixed to the static magnetic moment direction, the oscillating component of +Q mode at a fixed time rotates along the helical wave vector similarly to the static helical moments as illustrated in the bottom left panel of Fig. 1g. On the other hand, in the -Q mode, the rotating direction is opposite. The +Q and -Q modes are almost degenerate but the degeneracy may be lifted by the additional antiferromagnetic interaction\cite{kataoka}. 
As for the magnetic field dependence in the high field region above 400 Oe, the frequencies of observed two peaks decrease with $H_{\rm DC}$ below 1000 Oe. Above 1500 Oe, only one peak is observed, whose frequency rather increases with $H_{\rm DC}$. In Figs. 2 c and d, we show the magnetic susceptibility $\chi$ and the frequency of the observed peaks in the perpendicular configuration as functions of magnetic field.
The magnetic susceptibility is measured in magnetic fields perpendicular to the sample plane corresponding to the $H_{\rm DC} \perp H_{\rm AC}$ configuration.  
The metamagnetic transition from the conical magnetic state to the induced ferrimagnetic state is observed around 1400 Oe as a kink of magnetic susceptibility. As indicated by a thick gray bar, the change in the observed peak magnetic resonance mode(s) and its(their) field-dependence corresponds to this metamagnetic transition. 
The magnetic -field dependence of the mode frequency is in accord with the theoretical result\cite{kataoka}, thereby confirming the known character of the magnetic excitations in spin helix.

\begin{figure}[htbp]
\begin{center}
\includegraphics*[width=15cm]{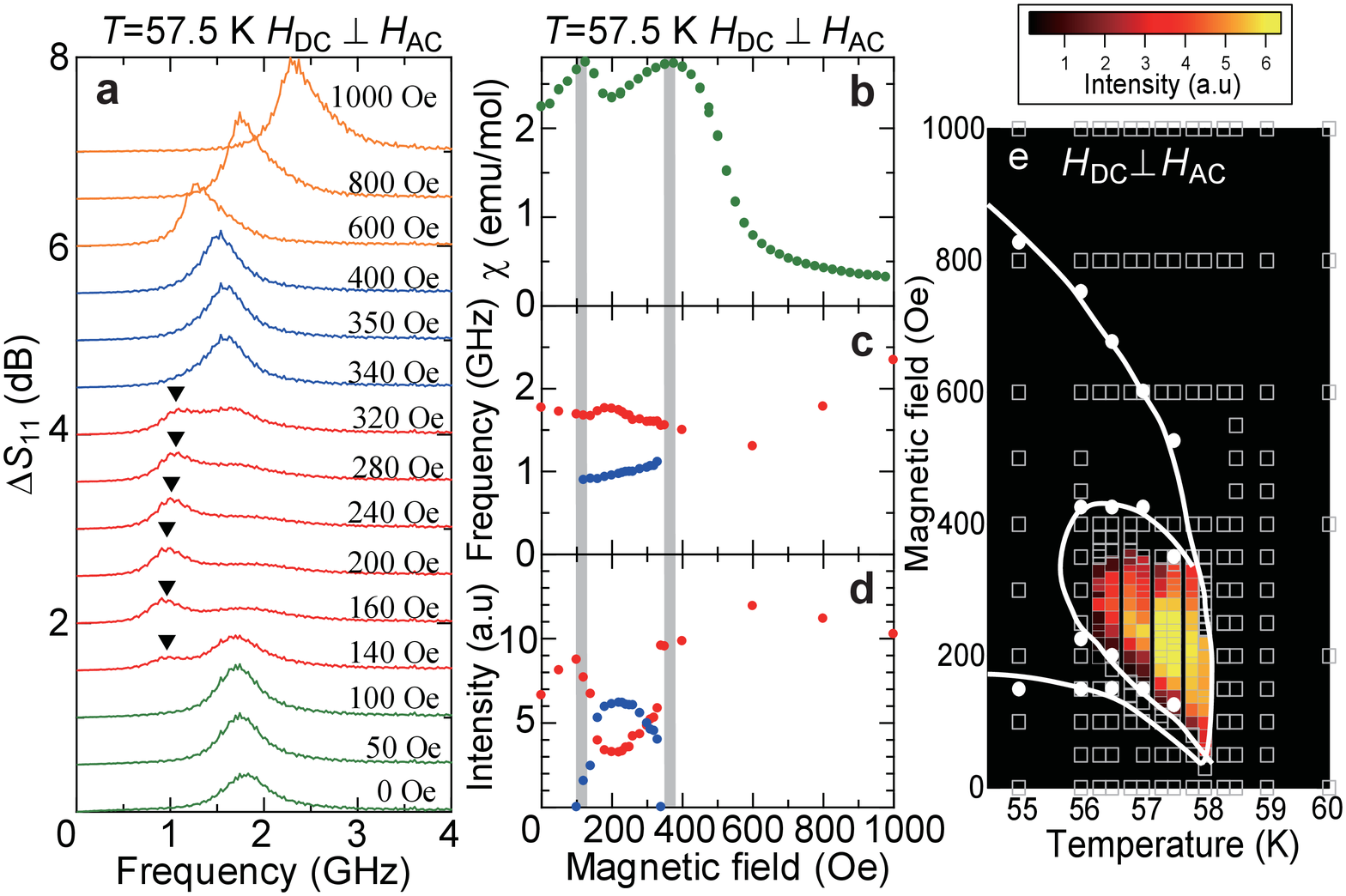}
\caption{{\bf Microwave response for $H_{\rm AC} \perp H_{\rm DC}$ in skyrmion crystal. a,} The microwave absorption spectra  $\Delta S_{11}$ for $H_{\rm DC} \perp H_{\rm AC}$ at various magnetic fields at 57.5 K. {\bf b,} The magnetic field variation of magnetic susceptibility at 57.5 K measured in the configuration corresponding to the $H_{\rm DC} \perp H_{\rm AC}$ microwave measurement. {\bf c,d,} The frequency ({\bf c}) and intensity ({\bf d}) of the magnetic modes in the microwave absorption spectra in a. {\bf e,} The intensity of the magnetic-resonance mode for $H_{\rm DC} \perp H_{\rm AC}$ (counter clock wise rotaional mode) in skyrmion crystal plotted in the $H-T$ phase diagram determined by the magnetic susceptibility measurement. Squares stand for the measured (temperatures, magnetic field) points.} 
\end{center}
\end{figure}

Next, we show the newly resolved magnetic resonances in the skyrmion. We have investigated the magnetic field dependence of microwave absorption spectra at 57.5 K to probe the magnetic excitations in skyrmion crystal (Scan 2 shown in Fig. 1a). In Fig. 3a,  we plot $\Delta S_{11}$ for $H_{\rm DC} \perp H_{\rm AC}$ at various magnetic fields. In the low-field helical magnetic state, one broad excitation is observed around 1.8 GHz. The splitting is not clearly observed in this temperature and the frequency is lower than that at 40 K. The peak is unchanged or slightly enhanced with $H_{\rm DC}$ up to 100 Oe. Nevertheless, the helical peak becomes quite weak and alternatively another peak emerges around 1 GHz above 140 Oe. Around 340 Oe, the low frequency peak disappears and the helical mode revives. In Figs. 3b-d, we show the frequency and intensity of the peaks as functions of magnetic field, comparing them with the magnetic susceptibility measured in this configuration. The susceptibility shows peaks around 120 Oe and 350 Oe, indicating the phase boundary between the helimagnetic  and skyrmion crystal states. The field range, where the low-frequency peak is observed, coincides with that for the skyrmion crystal phase. The intensity of the helical mode is not completely suppressed even in the skyrmion crystal phase. A similar coexistence of the helical and skyrmion crystal states is hinted by the neutron diffraction measurement for B20 compounds (the coexistence of the diffraction spot of $q || H$  due to the helical state and that of $q \perp H$ due to the skyrmion crystal)\cite{muhlbauer}. In the induced ferrimagnetic state above 600 Oe, the magnetic susceptibility is suppressed and the peak frequency readily increases with $H_{\rm DC}$. Figure 3e shows the contour mapping of the low-frequency peak intensity in the $H-T$ phase diagram around the skyrmion-crystal phase determined by the magnetic susceptibility measurement. It is clear from this figure that the low-frequency mode at 1.0 GHz is emergent only in the skyrmion crystal phase. Thus, the low-frequency peak is assigned to a magnetic excitation inherent in the skyrmion crystal. 

\begin{figure}[htbp]
\begin{center}
\includegraphics*[width=15cm]{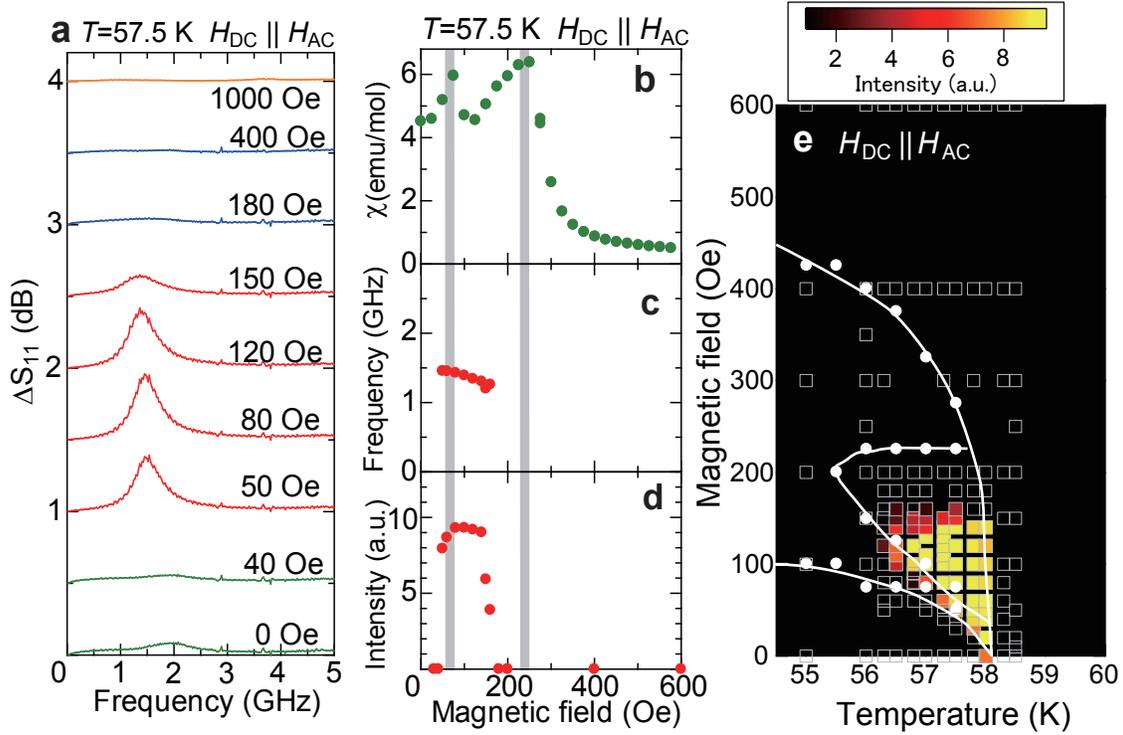}
\caption{{\bf Microwave response for $H_{\rm AC} || H_{\rm DC}$ in skyrmion crystal. a,} The magnetic field dependence of microwave absorption spectra  $\Delta S_{11}$ for $H_{\rm DC} || H_{\rm AC}$ at 57.5 K. {\bf b,} The magnetic field variation of magnetic susceptibility at 57.5 K measured in the configuration corresponding to the $H_{\rm DC} || H_{\rm AC}$ microwave measurement. {\bf c,d,} The frequency ({\bf c}) and intensity ({\bf d}) of the magnetic mode in the microwave absorption spectra for $H_{\rm DC} || H_{\rm AC}$. {\bf e,} The intensity of the magnetic-resonance mode for $H_{\rm DC} || H_{\rm AC}$ (breathing mode) in skyrmion crystal plotted in the $H-T$ phase diagram.  Squares stand for the measured (temperatures, magnetic field) points.} 
\end{center}
\end{figure}

Figure 4a shows the magnetic field variation of $\Delta S_{11}$ for $H_{\rm DC} || H_{\rm AC}$  at 57.5 K. The magnetic excitation cannot be excited by the AC magnetic field parallel to the magnetic moment. Note here that, conventional magnetic excitations are usually not observed in this configuration. Nevertheless, we discerned a peak around 1.5 GHz between 50 Oe and 150 Oe. In Fig. 4b-d, we plot the frequency and intensity of the peak as well as the magnetic susceptibility measured in this configuration. In the magnetic susceptibility, the sharp peaks are observed at 75 Oe and 250 Oe indicating the phase transition between the conical magnetic phase and skyrmion crystal phase. The reason why the magnetic field region of skyrmion-crystal phase is lower than that in the $H_{\rm DC} \perp H_{\rm AC}$ configuration is because of the demagnetization field as mentioned previously. The peak around 1.5 GHz is emergent around this magnetic field range. 
Also for this  $H_{\rm AC} || H_{\rm DC}$ configuration, we show the contour mapping of the peak intensity in the magnetic phase diagram determined by the magnetic susceptibility measurement. The region, where the absorption peak is emergent, almost coincides with the skyrmion crystal phase. (The small difference of the magnetic field ($\sim$ 50 Oe) is caused by the artifact perhaps due to the flux pining in the superconducting magnet.) The microwave absorption peak at 1.5 GHz in this parallel configuration can also be assigned to the excitation in the skyrmion crystal. 

Recently, Mochizuki numerically identified three spin-wave modes in  skyrmion crystal on the basis of Landau-Lifshitz-Gilbert equation\cite{mochi}. Two modes are counter-clockwise and clockwise rotational modes illustrated in Figs. 1d and e, in which the core of skyrmion rotates in the counter-clockwise and clockwise directions, respectively. These modes can be excited by the in-plane $H_{\rm AC}$. Petrova and Tchernyshyov analytically derived the similar rotational modes\cite{petrova}. The third mode is the breathing mode shown in Fig. 1f, in which the core of skyrmion expands and shrinks alternately and can be excited by the out-of-plane $H_{\rm AC}$. 
Since the skyrmion plane is always perpendicular to the DC magnetic field, the experimental configurations of $H_{\rm AC} \perp H_{\rm DC}$ and $H_{\rm AC} || H_{\rm DC}$ correspond to the in-plane and out-of-plane $H_{\rm AC}$, respectively. The observed mode at 1.5 GHz in the $H_{\rm AC} || H_{\rm DC}$ configuration (Fig. 4) can be undoubtedly assigned to the breathing mode of the skyrmion. For the in-plane polarization ($H_{\rm AC} \perp H_{\rm DC}$), on the other hand, only one skyrmion mode is experimentally observed, while two rotational modes are theoretically anticipated. 
It is, however, to be noted that the intensity and frequency of the counter-clockwise rotational mode are larger and lower, respectively, than those of the clockwise rotational mode according to the theoretical simulation.
The experimentally observed mode corresponds perhaps to the intense counter-clockwise rotational mode, while the 
higher lying clockwise mode may be mixed with the remaining helical mode and not clearly identified in the experiment.

In summary, we have observed the breathing and rotational modes of skyrmion 
in a helimagent Cu$_2$OSeO$_3$. The internal vibrations of skyrmion should be endowed with the magnetoelectric coupling. Therefore, nontrivial microwave responses such as directional dichroism are expected similarly to electromagnons in the THz regime\cite{youtarou}. In addition, these modes will be useful for the manipulation of skyrmions as already demonstrated by the numerical calculation\cite{mochi}.

\section*{Methods}
Single crystals of Cu$_2$OSeO$_3$ were grown by the chemical vapor transport\cite{seki,miller}. Magnetic susceptibility was measured in a SQUID magnetometer. The broad band microwave absorption spectra was measured in a superconducting magnet with use of a vector network analyzer. The detail of the microwave experiment is given in the Supplementary Information.

\section*{Acknowledgements}
The authors thank M. Mochizuki and N. Nagaosa for fruitful discussion. This research was supported in part by the Japan Society for the Promotion of Science (JSPS) through the \lq Funding Program for World-Leading Innovative R\&D on Science and Technology (First program)\rq , initiated by the Council for Science and Technology Policy (CSTP).

\section*{Author contributions}
Y. Onose and Y. Okamura carried out microwave measurement and analyzed data. S.I and S.S. carried
out crystal growth. The results were discussed and
interpreted by Y. Onose and Y. T.

\section*{Supplementary Information:Experimental setup of microwave absorption}
In order to observe magnetic excitations in skyrmion crystal, we have set up a broad band microwave measurement system with a vector network analyzer. The experimental setup is depicted in Fig. 5a. The semi rigid cable connected to the network analyzer is inserted into a superconducting magnet. A microstrip line$^1$ composed of narrow and broad copper plates is attached at the end of semi rigid cable (Fig. 5b). The other end of microstrip line is electrically shorted. Because the characteristic impedances of the semi rigid cable and microstrip line coincide with 50 $\Omega$, the microwave emitted from the network analyzer propagates along the semi rigid cable and microstrip line, reflects at the end of microstrip line, and then goes back to the network analyzer. The sample is inserted between the plates of microstrip line as shown in Figs. 5a and b. The AC magnetic field of microwave is polarized perpendicular to the propagation direction. The examples of observed reflection coefficient spectra $S_{11}$ are shown in Fig. 5c. The gradual decrease of background is caused by the loss of the semi rigid cable and microstrip line. The sharp dips in the 800 Oe spectrum are induced by the microwave absorption due to the magnetic excitations of sample. On the other hand, the frequency of magnetic excitations becomes higher than the measured frequency range at 5000 Oe. We have used the high field spectrum as the back ground and obtained the microwave absorption spectra $\Delta S_{11}$ by the subtraction in the unit of dB. As exemplified in Fig. 5d, smooth spectra are obtained using this procedure even without any numerical smoothing.\\
1. Pozar, D. M., Microwave Engineering (John Wiley \& Sons, Inc. 2005).

\begin{figure}[htbp]
\begin{center}
\includegraphics*[width=15cm]{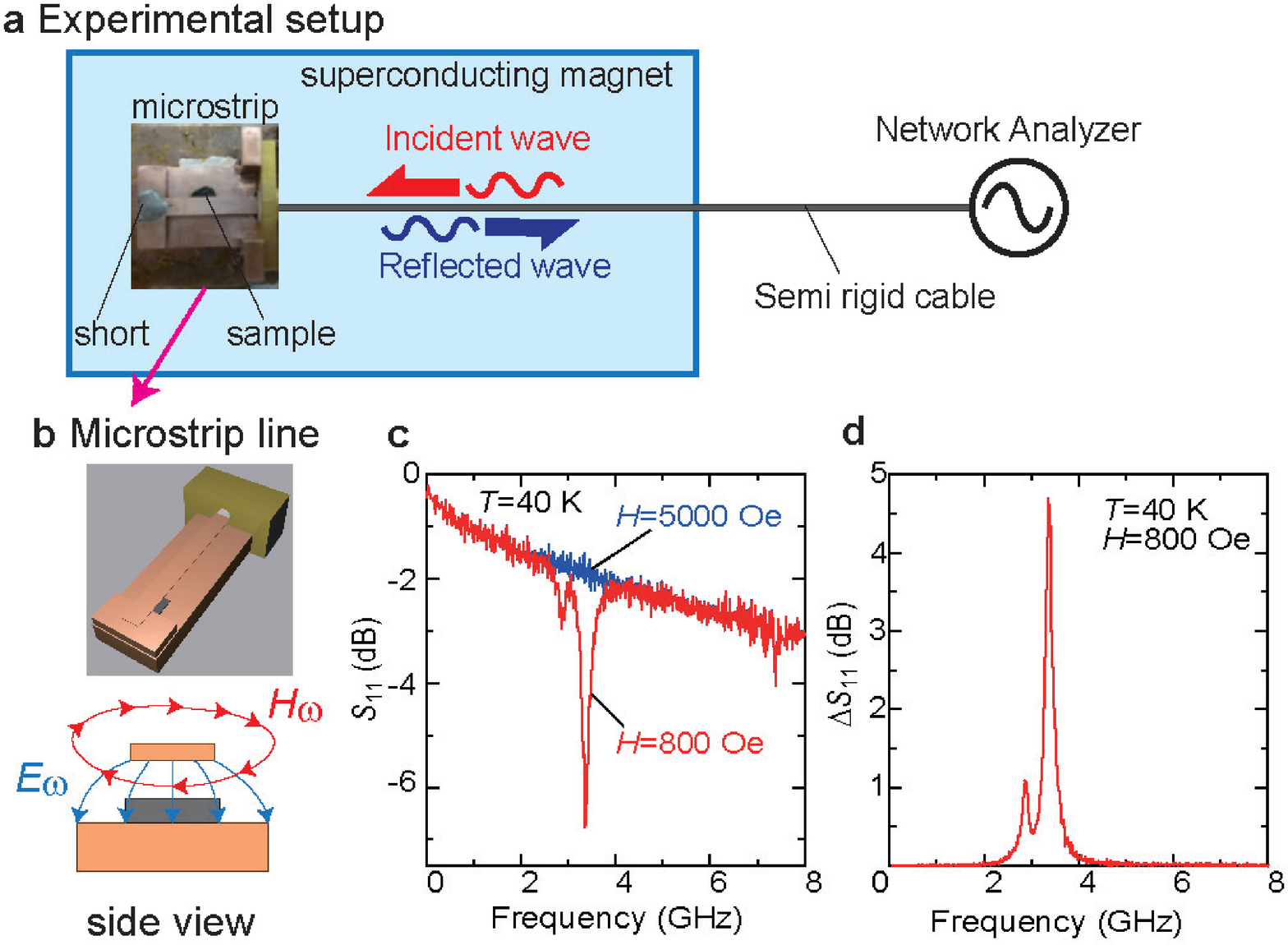}
\caption{{\bf Microwave absorption experiment a,b,} Illustrations of experimental setup and microstrip line. {\bf c,} Example of observed reflection coefficient spectra $S_{11}$. d, Obtained microwave absorption spectra $\Delta S_{11}$ by the subtraction. } 
\end{center}
\end{figure}

\begin{thebibliography}{}
\bibitem{skyrme} Skyrme, T. H. R. A unified theory of mesons and baryons. Nucl. Phys. 31, 556-559 (1962).
\bibitem{mlee} Lee, M., Kang, W., Onose, Y., Tokura, Y. \& Ong, N. P. Unusual Hall anomaly in MnSi under pressure. Phys. Rev. Lett. 102, 186601 (2009).
\bibitem{neubauer} Neubauer, A. {\it et al.} Topological Hall effect in the A phase of MnSi. Phys. Rev. Lett. {\bf 102,} 186602 (2009).
\bibitem{kanazawa} Kanazawa, N. {\it et al.} Large topological Hall effect in a short period helical magnet MnGe. Phys. Rev. Lett. {\bf 106,} 156603 (2011). 
\bibitem{muhlbauer} M\"{u}hlbauer, S. {\it et al.} Skyrmion lattice in a chiral magnet. Science {\bf 323,} 915-919 (2009).  
\bibitem{yu} Yu, X. Z. {\it et al.} Real-space observation of a two-dimensional skyrmion crystal. Nature {\bf 465,} 902-904 (2010).
\bibitem{yu2} Yu, X. Z. {\it et al.} Near room-temperature formation of a skyrmion crystal in thin-films of the helimagnet FeGe. Nature Material {\bf 10,} 106-109 (2011).
\bibitem{jonietz} Jonietz, F. {\it et al.} Spin transfer torques in MnSi at ultralow current densities. Science {\bf 330,} 1648-1651 (2010). 
\bibitem{bubble} Malozemoff, A. P. \& Slonczewski, J. C., {\it Magnetic domain walls in bubble materials} (Academic press 1979).
\bibitem{seki} Seki, S., Yu, X. Z., Ishiwata, S. \& Tokura, Y. Observation of skyrmions in a multiferroic material. Science {\bf 336,} 199-201 (2012).
\bibitem{maekawa} Maekawa, S. {\it Concepts in spin electronics} (Oxford science publications 2006).
\bibitem{schultz} Schulz, T. {\it et al.} Emergent electrodynamics of skyrmions in a chiral magnet arXiv:1202.1176 (2012).
\bibitem{zang} Zang, J., Mostovoy, M., Han, J. H. \& Nagaosa, N. Dynamics of skyrmion crystals in metallic thin films. Phys. Rev. Lett. {\bf 107,} 136804 (2011).  
\bibitem{osci} Gurevich, A.G. \& Melkov, G.A. {\it Magnetic Oscillations and Waves} (CRC press, 1996).
\bibitem{bos} Bos, J. -W. G., Colin, C. V. \& Palstra, T. T. M. Magnetoelectric coupling in the cubic ferrimagnet Cu$_2$OSeO$_3$. Phys. Rev. B {\bf 78} 094416 (2008).
\bibitem{belesi} Belesi, M. {\it et al.} Ferrimagnetism of the magnetoelectric compound Cu$_2$OSeO$_3$ probed by $^{77}$Se NMR. Phys. Rev. B {\bf 82,} 094422 (2010).
\bibitem{date} Date, M., Okuda, K. \& Kadowaki, K. Electron spin resonance in the itinerant-electron helical magnet MnSi. J. Phys. Soc. Jpn. {\bf 42,} 1555 (1977).
\bibitem{kataoka} Kataoka, M. Spin wave in systems with long period helical spin density waves due to the antisymmetric and symmetric interactions. J. Phys. Soc. Jpn. {\bf 56,} 3635-3647 (1987).
\bibitem{mochi} Mochizuki, M. Spin wave modes and their intense excitation effects in skyrmion crystals. Phys. Rev. Lett. {\bf 108,} 017601 (2012). 
\bibitem{petrova} Petrova, C. \& Tchernyshyov, O. Spin waves in a skyrmion crystal. Phys. Rev. B {\bf 84,} 214433 (2011). 
\bibitem{youtarou} Takahashi, Y., Shimano, R., Kaneko, Y., Murakawa, H. \& Tokura, Y. Magnetoelectric resonance with electromagnons in a perovskite helimagnet. Nature Phys. {\bf 8,} 121-125 
\bibitem{miller} Miller, K. H. {\it et al.} Magnetoelectric coupling of infrared phonons in single-crystal Cu$_2$OSeO$_3$ Phys. Rev. B {\bf 82,} 144107 (2010).  

\end{thebibliography}
\end{document}